%% file: main.tex
\documentclass[letterpaper]{article} 
\usepackage{aaai2027}
\usepackage[hyphens]{url}  
\usepackage{graphicx} 
\usepackage{natbib}  
\usepackage{caption} 
\usepackage{algorithm}
\usepackage{algorithmic}
\usepackage{placeins}
\usepackage{amsmath}
\usepackage{newfloat}
\usepackage{amssymb} 
\usepackage{listings}
\DeclareCaptionStyle{ruled}{labelfont=normalfont,labelsep=colon,strut=off} 
\floatstyle{ruled}
\newfloat{listing}{tb}{lst}{}
\floatname{listing}{Listing}

\usepackage{booktabs}

\title{\textit{HALO}: A P\textit{h}ysics-\textit{A}ware \textit{L}LM Agent Framework for Nanoph\textit{o}tonic Design}

\author{
    Yubo Zhang\textsuperscript{\rm 1}\equalcontrib,
    Jinlin Xiang\textsuperscript{\rm 1}\equalcontrib,
    Zijun Zhao\textsuperscript{\rm 1},
    Yang Zhao\textsuperscript{\rm 1},
    Eli Shlizerman\textsuperscript{\rm 1},
    Arka Majumdar\textsuperscript{\rm 1}
}

\affiliations{
    \textsuperscript{\rm 1}
    Department of Electrical \& Computer Engineering,
    University of Washington, Seattle, WA, USA
}

\begin{document}

\maketitle

\begin{abstract}
Language models have recently been applied to nanophotonic design,
including device generation, optimization, and tool-assisted simulation.
However, it remains less clear whether such systems can translate optical
objectives into complete simulation-ready designs, execute electromagnetic
analysis, and revise decisions from numerical feedback, or how these
capabilities should be evaluated reliably across diverse tasks. To address
these issues, we introduce HALO, a physics-aware framework that couples
language-model planners with a typed, machine-readable design specification,
electromagnetic simulation, diagnostic evaluation, and optional reuse of
prior failure trajectories in an iterative design loop. We further introduce
HALO-Bench, a 52-task benchmark spanning lab-derived, paper-derived, and
open-ended nanophotonic design tasks, to evaluate executable design
performance under a shared protocol. To study how workflow structure and
execution freedom affect scientific agent performance, we compare three
planner configurations: a Fixed Structured Workflow, an Autonomous Structured
Agent that retains the same simulation interface, and an Autonomous Coding
Agent that directly writes and executes simulation code. We find that the
Fixed Structured Workflow is consistently the most token-efficient and incurs
no observed code- or path-level failures, whereas autonomous coding can
achieve higher task success with stronger models at the cost of additional
operational failures. Because iterative design may repeatedly encounter
similar failure modes, we also study whether prior failed trajectories can
be reused to accelerate correction. On targeted cases requiring multiple
feedback rounds, retrieved failure feedback reduces the number of iterations
to first success and total token use. These findings clarify when scientific
agents benefit from explicit interfaces, autonomous execution, and reusable
design experience.
\end{abstract}

\begin{figure*}[t]
    \centering

  \includegraphics[
    width=0.90\textwidth,
    trim=0 17mm 0 17mm,
    clip
  ]{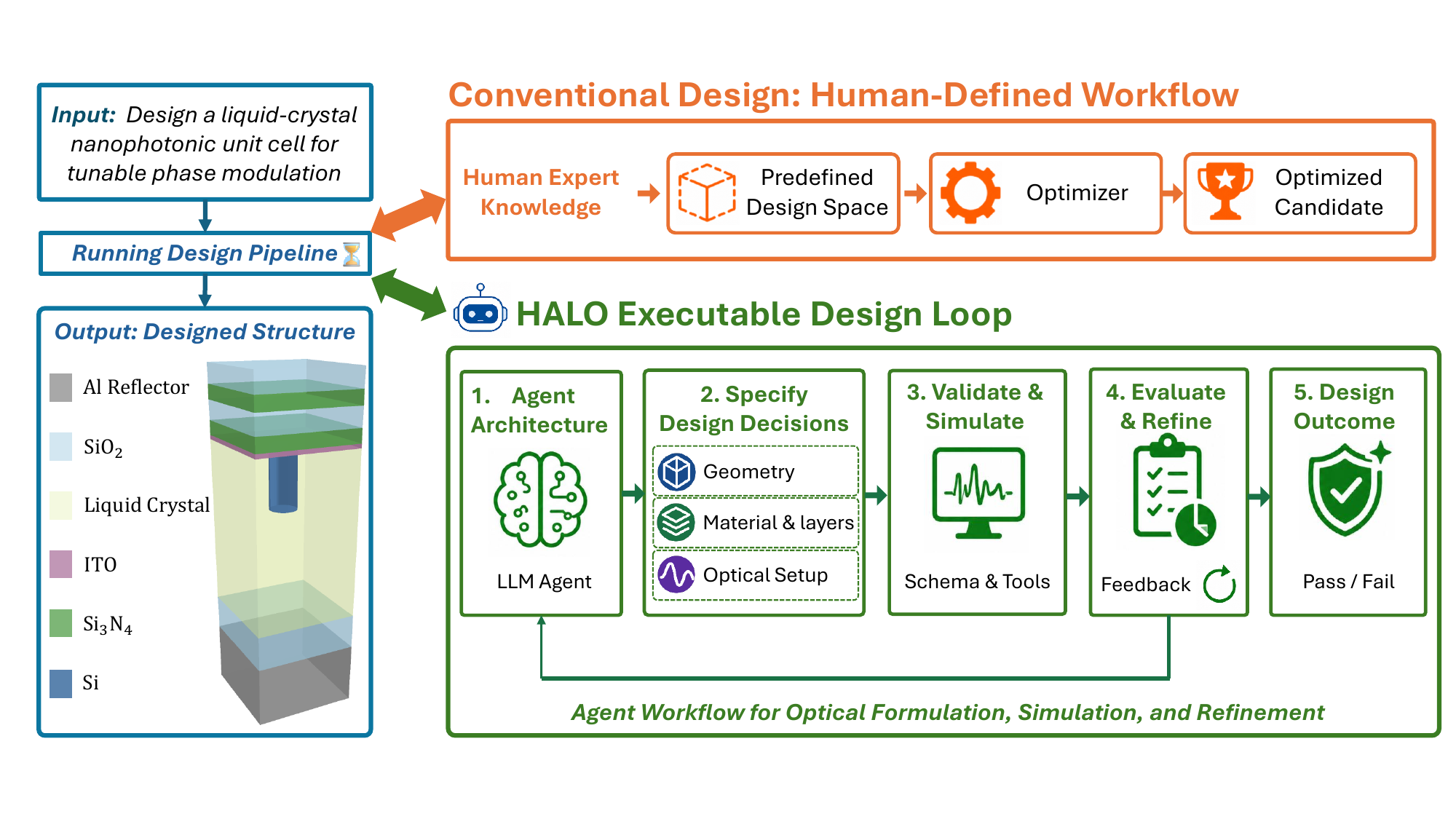}

    \caption{Overview of the HALO problem formulation. HALO converts an underspecified natural-language optical objective into explicit geometry, material assignments, simulation settings, and evaluation criteria, and iteratively revises the design using simulator feedback.}
    \label{fig:teaser}
\end{figure*}



\input{paper_section/1_Introduction}
\input{paper_section/2_Related_work}
\input{paper_section/3_Methods}
\input{paper_section/4_Experiments}

\input{paper_section/5_Discussion}

\bibliography{aaai2027}



\end{document}

%% file: paper_section/1_Introduction.tex
\section{Introduction}

Language-model systems increasingly plan, use tools, execute code, and
revise from feedback \cite{react,toolformer,autogen}. In scientific design, however, these general capabilities are not sufficient:
the system needs to translate an objective into an executable representation,
run the relevant tools, interpret numerical evidence, and improve within
limited evaluation budgets \cite{scienceagentbench,paperbench}.

Nanophotonic design makes this challenge concrete. A short optical goal
or device description may omit the geometry, material assignments,
illumination settings, solver configuration, design variables, and
evaluation metrics required for electromagnetic simulation. Classical
inverse-design methods are powerful after this representation and
objective have been specified \cite{molesky2018inverse}, but the
upstream conversion from natural-language intent to a simulation-ready
problem remains a substantial bottleneck.

We introduce \textbf{HALO}, a physics-aware executable framework that standardizes
the conversion from natural-language design intent into a simulation-ready
formulation and supports electromagnetic execution, diagnostic evaluation,
iterative revision, and optional reuse of prior failed trajectories. In this
work, physics awareness means that planner decisions are evaluated and
revised through electromagnetic simulation and task-specific physical
metrics. For auditability, HALO stores prompts, structured designs or
generated code, simulation outputs, scores, evaluator notes, and failure
records as design trajectories.

The associated \textbf{HALO-Bench} evaluates this process across \emph{Lab-Derived Simulation Tasks}, \emph{Paper-Derived Tasks}, and \emph{Open-Ended Tasks}. These regimes respectively test whether planners can recover known feasible designs, reconstruct simulation-ready devices from paper descriptions, and improve open-ended designs through iterative feedback.

Our central question is how much workflow and interface structure a scientific planner needs. We compare three planner configurations along two dimensions: workflow orchestration, ranging from fixed execution to autonomous agent control, and simulation access, ranging from a typed machine-readable interface to self-generated simulation code. We find that explicit scientific interfaces improve reliability and token efficiency while providing directly inspectable design records, whereas autonomous coding can achieve higher success with sufficiently capable models. Because some cases require multiple feedback rounds, we also examine whether retrieved failure feedback can accelerate convergence.

Our contributions are:
\begin{itemize}
\item \textbf{HALO}, a physics-aware framework for executable and
iterative nanophotonic-agent design and evaluation.
\item \textbf{HALO-Bench}, a 52-task benchmark spanning Lab-Derived
Simulation, Paper-Derived, and Open-Ended nanophotonic design tasks.
\item A system-level comparison of planner structure and autonomy, together with a targeted study showing that schema-assisted retrieval of failed
trajectories accelerates convergence on delayed-success cases.
\end{itemize}

%% file: paper_section/2_Related_work.tex
\section{Related Work}

\subsection{Nanophotonic Inverse Design}

Nanophotonic inverse design includes gradient-based optimization,
topology optimization, evolutionary search, parameter sweeps, and
fabrication-aware numerical methods
\cite{molesky2018inverse,jensen-sigmund-topopt,%
christiansen-sigmund-tutorial,piggott-demux,%
piggott-fabrication-constrained,hughes-nonlinear-adjoint,%
numerical-optimization-metasurfaces}. Machine-learning approaches learn
forward models, inverse mappings, generative representations, and
surrogate-assisted search strategies
\cite{tandem-nn,neural-adjoint,glonet,%
deep-learning-nanophotonics-review,%
ml-photonic-inverse-design-review,%
benchmarking-deep-learning-inverse-design,%
generative-multilayer-metasurfaces,%
diffusion-metasurface-inverse-design}.
These methods are effective once the device representation, design
variables, simulator, and objective have been specified. HALO addresses
a complementary problem: how a language-model planner formulates this
simulation-ready design problem from text and revises it using
electromagnetic feedback.

\subsection{LLM Agents for Scientific and Optical Design}

LLM agents support planning, tool use, code execution, and iterative
problem solving
\cite{react,toolformer,autogen,reflexion,metagpt,sweagent}.
Scientific-agent systems increasingly connect language models with
experiments, data analysis, simulation, and domain-specific software
\cite{scienceagentbench,chemcrow,coscientist,ai-scientist,%
scientific-computing-agent}. In optics and photonics, language models
have been applied to multilayer design, nanophotonic structure
generation, metasurface optimization, and tool-assisted simulation
\cite{optogpt,nanophotonic-llm-design,%
metachat-agentic-metasurface,autonomous-agentic-design-photonics,%
mcp-metaoptics-inverse-design,autonomous-metamaterial-agent,%
optiagent,chat-to-chip}.

Table~\ref{tab:related-comparison} compares representative systems by the role of the language model, execution control, explicit geometry and material representation, trajectory memory, and benchmark-based evaluation. HALO evaluates fixed and autonomous planner configurations within a shared executable design framework, while retaining complete trajectories for failure analysis. It also studies whether retrieving prior failed trajectories can accelerate convergence in later design attempts.

\input{Tables/Related_Work}

\subsection{Structured Spatial and Physical Reasoning}

Executable optical design requires jointly consistent decisions:
components must be arranged spatially, materials assigned to the intended
regions, solver conditions specified, and design variables connected to
measurable objectives. Spatial-reasoning benchmarks show that language
models can struggle to preserve geometric consistency in structured
representations
\cite{floorplanqa,geogrambench}. Related materials-science work studies
domain language models, scientific agents, and knowledge representations
that connect text, material information, and computational tools
\cite{matscibert,matsci-nlp,matkg,materials-experiment-kg,%
materials-knowledge-graph-mkg,mapps,agentic-materials-computation}.

These capabilities are often evaluated separately. HALO-Bench instead
tests how layout, material assignment, solver configuration, and design
variables interact within one executable workflow. Its structured
representation makes these decisions directly inspectable and supports
field-level evaluation before simulation feedback. This scope concerns
simulation-ready physical consistency rather than general materials
discovery.

%% file: Tables/Related_Work.tex
\begin{table*}[t]
\centering
\footnotesize
\setlength{\tabcolsep}{3pt}
\renewcommand{\arraystretch}{1.15}
\begin{tabular}{@{}p{2.5cm} p{1.7cm} p{2.2cm} c ccc ccc c@{}}
\toprule
& \multicolumn{3}{c}{System Architecture}
& \multicolumn{3}{c}{Geometry Grounding}
& \multicolumn{3}{c}{Material Grounding}
& \\
\cmidrule(lr){2-4}
\cmidrule(lr){5-7}
\cmidrule(lr){8-10}
System & LM Role & Control & Mem.
& Dimensionality & Layout & Valid.
& Rep. & Bind. & Valid.
& Bench. \\
\midrule
OptoGPT \cite{optogpt}
& inverse model
& single-pass
& $\times$
& 1D stack & \checkmark & $\times$
& \checkmark & \checkmark & $\times$
& $\times$ \\
Chat-to-Chip \cite{chat-to-chip}
& design generator
& single-pass
& $\times$
& 2D planar & \checkmark & \checkmark
& $\times$ & $\times$ & $\times$
& $\times$ \\
OPTIAGENT \cite{optiagent}
& learned policy
& optimizer-in-loop
& $\times$
& 2D lens & \checkmark & \checkmark
& \checkmark & \checkmark & \checkmark
& $\times$ \\
MetaChat \cite{metachat-agentic-metasurface}
& multi-agent system
& autonomous tool loop
& $\times$
& 2D freeform & \checkmark & $\times$
& \checkmark & \checkmark & $\times$
& $\times$ \\
Metamaterial Agent \cite{autonomous-metamaterial-agent}
& multi-agent system
& autonomous coding loop
& \checkmark
& 3D parameterized unit & $\times$ & $\times$
& $\times$ & $\times$ & $\times$
& $\times$ \\
\midrule
\textbf{HALO (ours)}
& planner configs
& fixed + autonomous
& \checkmark
& 3D typed geometry & \checkmark & \checkmark
& \checkmark & \checkmark & \checkmark
& \checkmark \\
\bottomrule
\end{tabular}
\caption{Representative language-model optical-design systems. Mem. indicates support for or explicit study of reusable memory. Rep. denotes an explicit machine-readable representation; Layout and Bind. indicate explicit spatial-layout and material-binding support, respectively; Valid. indicates validation of the corresponding representation or binding. Bench. denotes executable multi-task evaluation under a shared protocol.}

\label{tab:related-comparison}
\end{table*}

%% file: paper_section/3_Methods.tex
\section{Method}
\label{sec:method}

\subsection{HALO Workflow}
\label{sec:halo_workflow}

\begin{figure*}[t]
\centering
  \includegraphics[
    width=0.96\textwidth,
    trim=0 7mm 0 3mm,
    clip
  ]{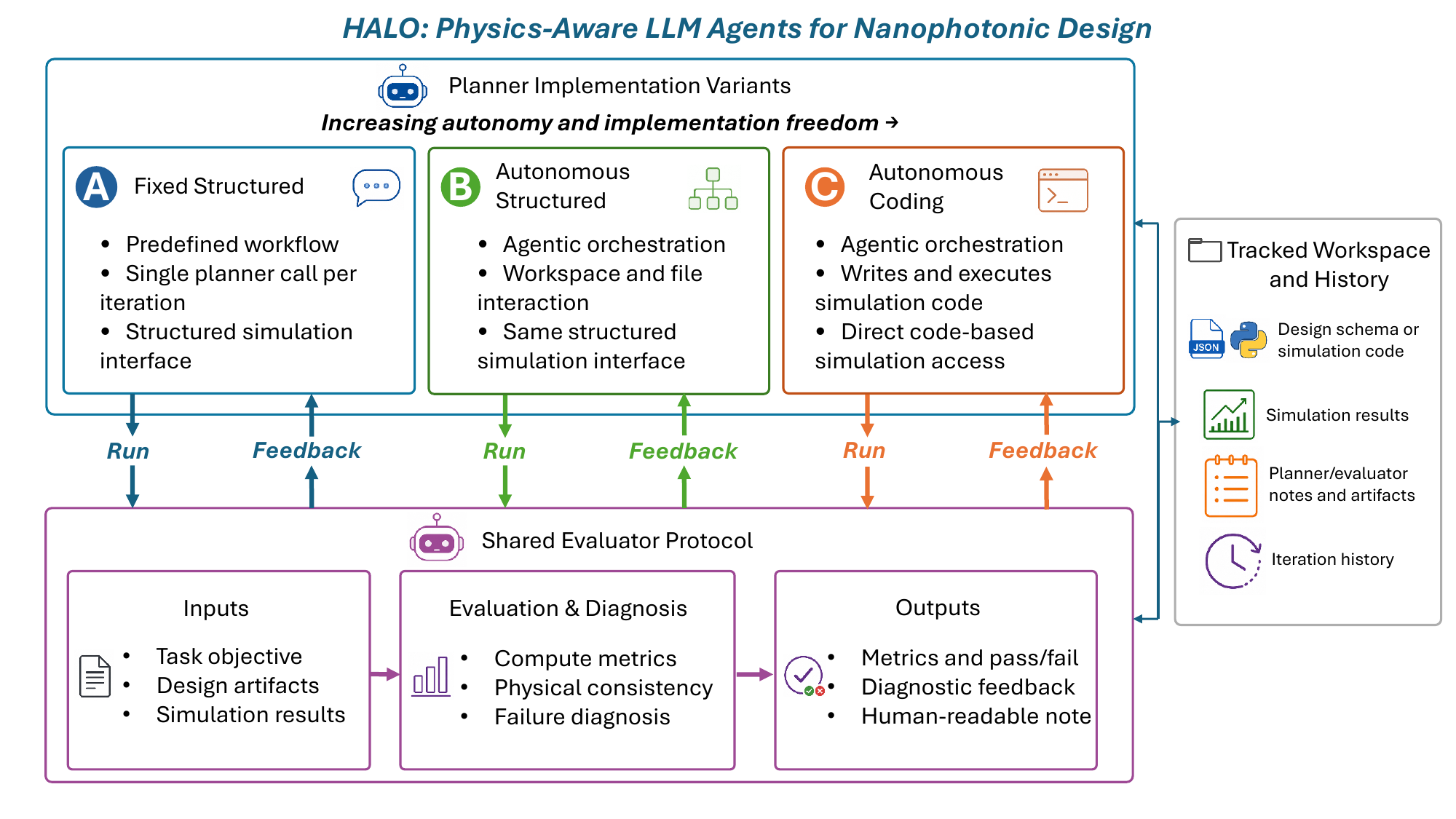}
\caption{HALO workflow. A planner converts a design goal into either a
design schema or executable simulation code, receives simulator-grounded
diagnostic feedback, and revises under a fixed outer-iteration budget.
HALO logs prompts, artifacts, outputs, scores, notes, and failures.}
\label{fig:method}
\end{figure*}

HALO is an executable framework that standardizes iterative
nanophotonic design as a
propose--execute--evaluate--revise loop. HALO represents each candidate
device using a typed, machine-readable simulation specification that
explicitly defines geometry and spatial relations, material assignments,
illumination and solver settings, design variables, and evaluation
objectives. We refer to this specification as the design schema.
Structured proposals are validated before RCWA execution; code-based
proposals are executed in a workspace and their outputs are collected
for the same evaluation protocol. Figure~\ref{fig:method} summarizes
the workflow.

HALO distinguishes two forms of history. Within-run trajectory history
contains previous iterations of the same case, including planner
outputs, schemas or generated code, simulation results, evaluator
feedback, and logs; it is available during the default closed-loop
experiments. Retrieval memory instead exposes prior design trajectories
from a memory pool and is enabled only in the memory ablation.

\subsection{HALO-Bench}
\label{sec:benchmark}

\begin{figure*}[t]
\centering
  \includegraphics[
    width=0.96\textwidth,
    trim=0 50mm 0 60mm,
    clip
  ]{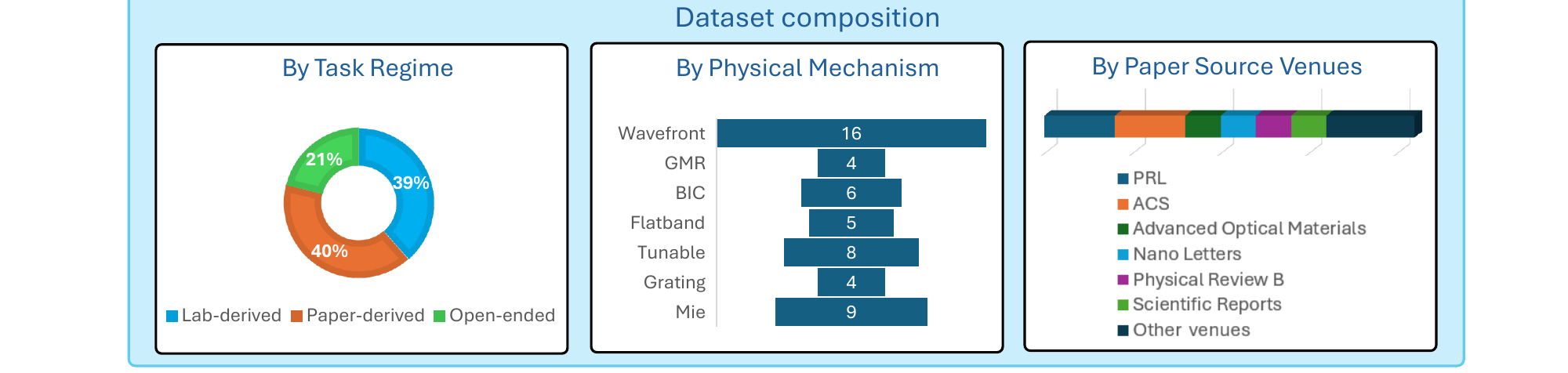}
\caption{Composition of HALO-Bench. The benchmark contains 20 Lab-Derived, 21 Paper-Derived,
and 11 Open-Ended tasks spanning seven nanophotonic mechanisms and multiple
publication venues.}
\label{fig:dataset}
\end{figure*}

HALO-Bench is the associated benchmark instantiated within HALO. It
contains three regimes, shown in Figure~\ref{fig:dataset}.

\paragraph{Lab-Derived Simulation Tasks.}
This setting contains 20 design problems adapted from prior projects in
our laboratory. Each task has a known workable reference configuration
previously verified using the simulation workflow employed in this study
or a compatible workflow. These cases provide feasible simulation
targets, but they should not be interpreted as experimentally fabricated
or measured devices.

\paragraph{Paper-Derived Tasks.}
This setting contains 21 short natural-language design goals generated
from published nanophotonic device descriptions through LLM-assisted
drafting followed by manual refinement. The prompts describe the
relevant structure and optical objective without giving the evaluated
planner the complete source paper. Here, recovering a Paper-Derived
design means producing a working simulation-compatible design that
captures the reported structure and target behavior, not reproducing
every experimental curve, fabrication detail, or reported numerical
value.

\paragraph{Open-Ended Tasks.}
This setting contains 11 tasks derived from a selected subset of
Paper-Derived Tasks by relaxing selected numerical or structural
constraints while preserving the broad physical objective. They require
exploration and iterative improvement rather than exact reconstruction.
Detailed generation prompts and source-case mappings are provided in the
supplementary material.

\subsection{Planner Configurations}
\label{sec:agent_interfaces}

The three planner configurations form a progression of increasing autonomy while sharing the same task inputs, outer-iteration budget, wall-clock budget, and evaluator protocol. Fixed Structured Workflow uses a predefined workflow and HALO's structured simulation interface. Autonomous Structured Agent adds agentic orchestration and workspace interaction while retaining the same structured simulation interface. Autonomous Coding Agent retains autonomous orchestration but additionally allows the agent to write and execute simulation code directly, removing the predefined interface constraint. Accordingly, Fixed Structured Workflow versus Autonomous Structured Agent provides a practical comparison of fixed versus agentic orchestration under the same structured simulation interface, whereas Autonomous Structured Agent versus Autonomous Coding Agent compares
structured versus code-based simulation access under the same
autonomous workspace.

\paragraph{Fixed Structured Workflow.}
This configuration is implemented as a fixed LangGraph workflow. The
execution path is predefined, with no dynamic tool-selection branch; the
only conditional branch occurs at the end of an outer design iteration,
where the run stops if the evaluator confirms success and otherwise
continues. Each outer design iteration contains one planner LLM call.
The planner returns a structured design output requested through
prompting and checked afterward with Pydantic validation. The workflow
then invokes HALO's predefined validation and RCWA simulation tools. The
planner cannot freely read or write files, change the workflow, write
simulator code, or select arbitrary tools.

\paragraph{Autonomous Structured Agent.}
This configuration is implemented with a LangChain Deep Agent using a
local-shell/filesystem workspace. It can perform multistep analysis, read
and write files, maintain notes, inspect artifacts, and decide
intermediate actions. Its final design representation remains
structured, and all formal RCWA simulations must use HALO's predefined
structured simulation interface. It is therefore more autonomous in
orchestration than the fixed workflow while retaining the same
scientific execution boundary.

\paragraph{Autonomous Coding Agent.}
This configuration uses the same autonomous workspace but directly
writes, executes, inspects, and revises Python simulation code using the
available RCWA package. It is not constrained to HALO's predefined
design schema for simulation access. This provides greater
implementation freedom, but the physical design decisions are less
structurally auditable even though files, code, commands, outputs, and
logs are retained. It also introduces code, API, path, artifact, and
output-format failure modes.

\subsection{Common Evaluation Protocol}
\label{sec:evaluation_protocol}

All planner configurations follow a shared evaluator-generation protocol with the same task objective, metric requirements, execution environment, and pass/fail output contract. Because schema-based and code-based planners produce heterogeneous artifacts, the evaluator inspects the available design artifact and simulation output and generates a case-specific executable evaluation script. The generated script converts saved outputs into numerical metrics and a score/pass decision, and is then fixed and reused across subsequent iterations of that run. The evaluator then generates a human-readable diagnostic note that summarizes the numerical result and explains the reason for success or failure; these notes were independently sanity-checked by additional optics experts for physical and logical consistency with the corresponding design objectives and numerical results.
Because evaluator generation uses the same underlying model as the
corresponding planner, the reported results should be interpreted as
system-level comparisons of complete model--configuration pipelines
under a shared evaluator-generation protocol, rather than as isolated
measurements of planner quality.

\subsection{Optional Retrieval Memory}
\label{sec:retrieval_memory}

Retrieval memory stores prior failed trajectories that may contain
useful geometry choices, parameter ranges, evaluation signals, and
recovery strategies. The memory ablation compares no retrieved memory,
schema-assisted retrieval, and a same-task failure-memory control.
Unlike within-run history, retrieval memory can introduce information
from prior cases and is analyzed separately from the default
closed-loop success comparison.

%% file: paper_section/4_Experiments.tex
\section{Experiments}
\label{sec:experiments}

Our experiments ask: (i) whether models can produce an initially
plausible structured design before feedback; (ii) how planner structure
affects closed-loop success and token efficiency; (iii) whether
retrieved failed trajectories accelerate delayed-success cases; and
(iv) where each configuration fails.

\subsection{Experimental Setup}
\label{sec:experimental_setup}

HALO-Bench contains 52 tasks: 20 Lab-Derived Simulation Tasks, 21
Paper-Derived Tasks, and 11 Open-Ended Tasks. Closed-loop runs compare
the Fixed Structured Workflow, Autonomous Structured Agent, and
Autonomous Coding Agent. Each model--configuration--case is
run once with the same task prompt, task-specific objective, common
evaluator protocol, at most five outer design iterations, and a
one-hour wall-clock limit. Runs stop early after confirmed success. The
one-hour limit prevents individual cases from blocking the full
benchmark and defines a practical bounded evaluation setting.

Experiments were run on one workstation with an NVIDIA RTX 6000 GPU. For each model–configuration pair, the planner and evaluator use the same underlying language model. We record planner and evaluator token usage
separately. 

Because the three planner configurations perform different amounts of tool use, code execution, and debugging, their internal workloads are not directly comparable. We therefore report successful cases per million recorded language-model tokens as a measure of token efficiency. This metric is most reliable for comparisons within the same model family, since token accounting differs across providers. Each case is run once because of the scale and computational cost of the benchmark.

\input{Tables/Spatial_Material}
\input{Tables/Success_Rate}
\input{Tables/Memory_Ablation}

\subsection{Structured Design Understanding}
\label{sec:understanding}

Closed-loop performance combines prompt understanding, artifact
construction, simulation, feedback interpretation, and tool execution.
We isolate first-step design formulation using the 20 Lab-Derived Simulation Tasks, which provide paired natural-language descriptions and human-annotated, simulation-ready reference schemas. Evaluating only the
original prompts would be insufficiently diagnostic because they already
describe known feasible configurations. We therefore construct controlled
prompt--schema perturbations by modifying selected requirements in the
natural-language description together with their corresponding fields in
the reference schema. Numerical values and supported material or
categorical substitutions are updated synchronously, preserving a
consistent input--reference pair while testing whether the model follows
the modified requirement rather than reproducing the original design.

The model receives only the input prompt for each condition, while the paired schema is used solely as the evaluation reference. Each task is evaluated under the original prompt and three controlled perturbation settings, yielding 80 prompt–schema pairs per model. Exact perturbation rules, conditions, and sample counts are provided in the supplementary material.

Before receiving validation, simulation results, or iterative feedback, the model generates a design schema, which is then evaluated against the paired human-annotated reference using the same fixed procedure. We evaluate layout representation, numerical binding, material assignment, and sweep-variable specification. Detailed metric definitions and scoring rules are provided in the supplementary material. We restrict this field-level comparison to the Lab-Derived tasks because the Paper-Derived and Open-Ended tasks may admit multiple simulation-compatible formulations rather than a unique reference schema.

Table~\ref{tab:understanding} shows more than 93\% overall accuracy for all evaluated models on the perturbed Lab-Derived tasks. Thus, closed-loop failures cannot be explained solely by an inability to produce an initially plausible structured design. Reliability also depends on preserving, executing, and revising these decisions from simulation feedback.

\subsection{Closed-Loop Success and Efficiency}
\label{sec:closed_loop}

Table~\ref{tab:end_to_end_success_tokens} reports complete
propose--execute--evaluate--revise trajectories over all three
HALO-Bench regimes.

Observed success varies with both the underlying model and planner
implementation. The
highest observed macro-average success is 79.8\%, achieved by the
Autonomous Coding Agent with Claude Sonnet 4.6. Autonomous Coding Agent also
improves substantially across Gemini models, from 32.3\% with Gemini
2.5 Flash to 68.8\% with Gemini 3.1 Pro Preview, consistent with stronger models being better able to exploit
code-generation freedom.

Autonomy is not uniformly beneficial. With Gemini 2.5 Flash and Gemini
2.5 Pro, the Fixed Structured Workflow achieves higher
macro-average success than either autonomous configuration. The
Autonomous Structured Agent also does not consistently
outperform the fixed workflow. Across every model block, the Fixed Structured Workflow
achieves the highest number of successful cases per million
recorded tokens; with Claude Sonnet 4.6, it achieves 2.327
successes per million tokens, compared with 0.474 for
Autonomous Structured Agent and 0.604 for Autonomous Coding
Agent. These comparisons apply
under five outer design iterations, a one-hour wall-clock limit, and
non-equalized internal tool actions.

\subsection{Memory Retrieval Ablation}
\label{sec:memory_retrieval_ablation}

Cases that succeed only after multiple feedback rounds motivate a targeted study of whether retrieved failure history can accelerate convergence. We first examine how to retrieve relevant failed trajectories from memory, comparing text-only retrieval with a schema-assisted alternative. Because text-only matching may miss physically related trajectories when descriptions differ in wording or order, our schema-assisted method augments text similarity with schema-level similarity over geometry, material assignments, optical setup, and sweep variables. The text- and schema-based similarity scores are combined into a final ranking score, and the three highest-ranked failed trajectories are retrieved. Detailed retrieval design, scoring, and the comparison with text-only retrieval are provided in the supplementary material.

We conduct the ablation on 16 cases in which the No Retrieved Memory baseline eventually succeeded but required more than one iteration. Using the same Gemini 3.1 Pro Preview Autonomous Structured Agent, we compare No Retrieved Memory, Schema-Assisted Top-3 Retrieval, and Same-Task Failure Memory (Top-1 Control). The retrieval pool is constructed from failed trajectories produced by prior Gemini 3.1 Pro Preview runs; successful trajectories and success-revealing artifacts are excluded. We allow task-level overlap because the experiment evaluates convergence acceleration rather than cross-task memory generalization. Accordingly, this ablation evaluates retrieval-assisted reuse within
an available failure-history pool, rather than transfer to previously
unseen tasks. Schema-Assisted Top-3 ranks all eligible failed trajectories using the retrieval score. The Same-Task Top-1 control bypasses retrieval and directly provides one failed trajectory from the corresponding task.

Among the 16 baseline-delayed-success cases, Schema-Assisted Top-3 Retrieval converts 10 cases from multi-iteration to first-iteration success. It reduces the mean first-success iteration from 2.44 to 1.44, lowers the median per-case token usage from 1.67M to 0.34M, and reduces aggregate token use from 41.38M to 14.79M, a 64.3\% reduction. The Same-Task Top-1 Control also accelerates convergence, reducing the mean first-success iteration to 1.75 and aggregate token use to 20.71M, a 50.0\% reduction relative to No Retrieved Memory. This condition serves as a diagnostic control because it directly provides a failed trajectory from the corresponding task; it should not be interpreted as either a realistic first-time deployment setting or a strict upper bound.

\subsection{Failure Analysis}
\label{sec:failure_analysis}

Each failed case is assigned one dominant category using a consistent annotation rule based on the complete trajectory and final artifacts; category definitions and representative examples are provided in the supplementary material. Successful cases are
excluded; if a later valid iteration succeeds, earlier failed iterations
do not count as a case failure.

\input{Tables/Failure_Analysis}

Table~\ref{tab:failure_analysis} aggregates 100 Fixed Structured Workflow
failures, 133 Autonomous Structured Agent failures, and 115 Autonomous Coding Agent
failures over all five evaluated models. Fixed Structured Workflow failures
occur mainly at simulator/evaluation or scientific-target stages and
show no code/path errors. Autonomous Structured Agent has more missing or
invalid results and setup/sweep-management errors. Autonomous Coding Agent
adds code/tool/API and artifact/path failures. This operational burden
does not make Autonomous Coding Agent uniformly worse: it achieves the highest
observed success with a strong model, but pays a larger software and
artifact-management cost.

%% file: Tables/Spatial_Material.tex
\begin{table}[t]
\centering

\resizebox{\columnwidth}{!}{
\begin{tabular}{lccccc}
\toprule
Model &
Layout &
Numeric &
Material &
Sweep &
Overall \\
\midrule
Gemini 2.5 Flash
& 95.9 & 90.6 & 92.9 & 93.3 & 93.2 \\

Gemini 2.5 Pro
& 96.9 & \textbf{97.1} & \textbf{97.9} & 96.8 & \textbf{97.2} \\

Gemini 3.1 Pro Preview
& \textbf{97.6} & 93.2 & 92.7 & \textbf{97.7} & 95.3 \\
\midrule
Claude Haiku 4.5
& 94.2 & \textbf{94.7} & 97.1 & 97.3 & 95.8 \\

Claude Sonnet 4.6
& 97.4 & 91.6 & 94.4 & 97.7 & 95.3 \\

Claude Opus 4.8
& \textbf{99.6} & 94.6 & \textbf{98.3} & \textbf{98.3} & \textbf{97.7} \\
\bottomrule
\end{tabular}}
\caption{First-step structured design accuracy (\%) on perturbed Lab-Derived Simulation Tasks. Each model generates a schema without simulation feedback or iterative revision. Predictions are compared against human-annotated reference schemas using a fixed evaluation procedure. Results are averaged over 80 prompt–schema pairs per model. Metric definitions and scoring details are provided in the supplementary material. Overall is the unweighted mean of the four reported dimensions. Bold values indicate the best result within each model family.}
\label{tab:understanding}
\end{table}

%% file: Tables/Success_Rate.tex
\begin{table*}[t]
\centering
\scriptsize
\setlength{\tabcolsep}{3pt}

\begin{tabular}{llrrrrrrrr}
\toprule
Model & Configuration &
Lab (\%) &
Paper (\%) &
Open (\%) &
Macro avg. (\%) &
Plan tok. &
Eval tok. &
Total tok. &
Succ./M tok. $\uparrow$ \\
\midrule

Gemini 2.5 Flash
& Fixed Structured Workflow
& \textbf{90.0} & \textbf{23.8} & \textbf{36.4}
& \textbf{50.1} & \textbf{49k} & 573k & \textbf{622k}
& \textbf{0.835} \\

& Autonomous Structured Agent
& 30.0 & 4.8 & 0.0
& 11.6 & 1.49M & 308k & 1.80M
& 0.075 \\

& Autonomous Coding Agent
& 55.0 & \textbf{23.8} & 18.2
& 32.3 & 1.28M & \textbf{227k} & 1.50M
& 0.231 \\

\midrule

Gemini 2.5 Pro
& Fixed Structured Workflow
& \textbf{75.0} & \textbf{42.9} & \textbf{63.6}
& \textbf{60.5} & \textbf{52k} & 879k & \textbf{932k}
& \textbf{0.640} \\

& Autonomous Structured Agent
& 50.0 & \textbf{42.9} & 18.2
& 37.0 & 1.76M & 587k & 2.35M
& 0.172 \\

& Autonomous Coding Agent
& 55.0 & 19.0 & 27.3
& 33.8 & 3.53M & \textbf{235k} & 3.76M
& 0.092 \\

\midrule

Gemini 3.1 Pro Preview
& Fixed Structured Workflow
& 85.0 & 47.6 & \textbf{63.6}
& 65.4 & \textbf{59k} & 773k & \textbf{833k}
& \textbf{0.785} \\

& Autonomous Structured Agent
& 90.0 & \textbf{52.4} & 54.5
& 65.6 & 2.01M & 406k & 2.42M
& 0.278 \\

& Autonomous Coding Agent
& \textbf{95.0} & 47.6 & \textbf{63.6}
& \textbf{68.8} & 1.44M & \textbf{263k} & 1.71M
& 0.405 \\

\midrule

Claude Sonnet 4.6
& Fixed Structured Workflow
& \textbf{95.0} & 66.7 & 45.5
& 69.1 & \textbf{22k} & 292k & \textbf{314k}
& \textbf{2.327} \\

& Autonomous Structured Agent
& \textbf{95.0} & 57.1 & 54.5
& 68.9 & 1.17M & 328k & 1.50M
& 0.474 \\

& Autonomous Coding Agent
& 90.0 & \textbf{85.7} & \textbf{63.6}
& \textbf{79.8} & 1.18M & \textbf{181k} & 1.37M
& 0.604 \\

\midrule

Claude Haiku 4.5
& Fixed Structured Workflow
& 75.0 & \textbf{47.6} & \textbf{45.5}
& \textbf{56.0} & \textbf{32k} & 981k & \textbf{1.01M}
& \textbf{0.571} \\

& Autonomous Structured Agent
& 75.0 & 38.1 & 36.4
& 49.8 & 749k & 1.16M & 1.91M
& 0.272 \\

& Autonomous Coding Agent
& \textbf{80.0} & \textbf{47.6} & 36.4
& 54.7 & 1.85M & \textbf{471k} & 2.32M
& 0.249 \\

\bottomrule
\end{tabular}

\caption{
Closed-loop success and token use. Lab, Paper, and Open denote
Lab-Derived Simulation, Paper-Derived, and Open-Ended Tasks.
Macro avg. is the simple average of the Lab, Paper, and Open success
rates. Plan, Eval, and Total tok. report mean per-run token counts over runs.
Succ./M tok. is the number of successful runs per million
aggregate recorded tokens.
Bold marks the best value within each model block.
}
\label{tab:end_to_end_success_tokens}

\end{table*}

%% file: Tables/Memory_Ablation.tex
\begin{table*}[t]
\centering
\small
\renewcommand{\arraystretch}{1.18}
\setlength{\tabcolsep}{5pt}

\begin{tabular*}{\textwidth}{
@{\extracolsep{\fill}}
l
c
c
c
c
c
}
\toprule
Memory condition
& Succ.@1
& Mean iter. $\downarrow$
& Median tokens (M) $[Q_1,Q_3]$ $\downarrow$
& Aggregate tokens (M) $\downarrow$
& Succ./M tokens $\uparrow$ \\
\midrule

No Retrieved Memory
& 0/16
& 2.44
& 1.67 [1.09, 2.88]
& 41.38
& 0.387 \\

Schema-Assisted Top-3 Retrieval
& \textbf{10/16}
& \textbf{1.44}
& \textbf{0.34 [0.20, 1.17]}
& \textbf{14.79 ($-$64.3\%)}
& \textbf{1.082} \\

Same-Task Top-1 Control
& 8/16
& 1.75
& 0.47 [0.26, 1.81]
& 20.71 ($-$50.0\%)
& 0.773 \\

\bottomrule
\end{tabular*}
\caption{Targeted trajectory-memory ablation on 16 baseline-delayed
success cases. The subset is selected such that the No Retrieved Memory condition succeeds after, but not at, the first iteration; therefore, its Success@1 is zero by construction. All conditions eventually succeed; metrics report
convergence speed and language-model token efficiency. The same-task
Top-1 control provides one prior failed trajectory from the corresponding
task.}
\label{tab:memory-ablation}
\end{table*}

%% file: Tables/Failure_Analysis.tex
\begin{table}[t]
\centering
\scriptsize
\setlength{\tabcolsep}{3pt}

\begin{tabular}{lrrr}
\toprule
Failure category & \shortstack{Fixed\\Structured\\Workflow} &
\shortstack{Autonomous\\Structured\\Agent} &
\shortstack{Autonomous\\Coding\\Agent} \\
\midrule
Timeout / incomplete evaluation
& 29 (29.0\%) & 34 (25.6\%) & 33 (28.7\%) \\

Generated artifact / path error
& 0 (0.0\%) & 1 (0.8\%) & 21 (18.3\%) \\

Code / tool / API error
& 0 (0.0\%) & 0 (0.0\%) & 29 (25.2\%) \\

Simulator / runtime error
& 34 (34.0\%) & 16 (12.0\%) & 13 (11.3\%) \\

Missing / invalid result
& 4 (4.0\%) & 38 (28.6\%) & 13 (11.3\%) \\

Wrong setup / sweep
& 9 (9.0\%) & 25 (18.8\%) & 0 (0.0\%) \\

Target not met
& 24 (24.0\%) & 19 (14.3\%) & 6 (5.2\%) \\
\midrule
Total failures
& 100 & 133 & 115 \\
\bottomrule
\end{tabular}
\caption{Failure composition aggregated over all five evaluated models.
Each entry reports the number of failed cases, with percentages computed
within the failed cases of each planner configuration. Successful cases
are excluded, and each failed case is assigned one dominant category. Labels describe the dominant terminal failure of each case rather than every error observed during its trajectory.}
\label{tab:failure_analysis}
\end{table}

%% file: paper_section/5_Discussion.tex
\section{Discussion and Conclusion}
\label{sec:discussion}

HALO frames nanophotonic-agent design as an executable loop from
natural-language intent to simulation-ready formulation, simulator execution,
diagnostic evaluation, revision, and optional reuse of prior failures.
Across HALO-Bench, the Fixed Structured Workflow is the most
token-efficient, avoids the code- and path-level failures observed in
Autonomous Coding, and produces directly inspectable design records.
Autonomous Coding can achieve the highest success with a strong model,
but incurs additional software, artifact, and output-format failures.
The Autonomous Structured Agent further suggests that added orchestration
is not uniformly beneficial when the structured simulation boundary is retained.

Memory provides a second lever. On cases solved by the baseline only after
multiple iterations, retrieved failure feedback reduced both convergence
time and total token use, helping already-solvable cases converge more efficiently.

\paragraph{Limitations.}
HALO-Bench primarily covers periodic nanophotonic structures, and each
model--configuration--case is evaluated once. The memory study targets
baseline-delayed-success cases and measures convergence rather than
benchmark-wide success gains. Final designs may also require
application-specific, higher-fidelity validation.